\documentclass[aip,jcp,reprint]{revtex4-2}
\usepackage{chemformula} 
\usepackage[T1]{fontenc} % Use modern font encodings
\usepackage{comment}
\def\lang{\langle}
\def\rang{\rangle}

\begin{document}

\title{Coupled-Cluster Imaginary-Time Evolution and the Coupled-Cluster Energy Variance}

\author{Yuhang Ai}
\email{yai@caltech.edu}
\affiliation{Division of Chemistry and Chemical Engineering, California Institute of Technology,
Pasadena, California 91125, USA}
\affiliation{Marcus Center for Theoretical Chemistry, California
Institute of Technology, Pasadena, California 91125, USA}
\author{Huanchen Zhai}
\affiliation{Division of Chemistry and Chemical Engineering, California Institute of Technology,
Pasadena, California 91125, USA}
\author{Garnet Kin-Lic Chan}
\email{gkc1000@gmail.com}
\affiliation{Division of Chemistry and Chemical Engineering, California Institute of Technology,
Pasadena, California 91125, USA}
\affiliation{Marcus Center for Theoretical Chemistry, California
Institute of Technology, Pasadena, California 91125, USA}

\begin{abstract}
    We discuss a coupled-cluster formalism for carrying out imaginary-time evolution from an arbitrary reference, and study the properties of the resulting evolution trajectories. The evolution converges to a solution of the standard coupled-cluster amplitude equations in the long-time limit if a finite valued limit exists, but when such a limit does not exist, the trajectories still contain additional information beyond the standard solutions.  We introduce the coupled-cluster energy variance which through its minima identifies physically regularized coupled-cluster amplitudes when the solutions of the amplitude equations are unreasonable. We demonstrate the value of this formalism in several exploratory examples within single- and multi-reference coupled-cluster formulations. 
\end{abstract}

\maketitle

\section{Introduction}

Imaginary-time evolution (ITE) is a powerful algorithm for computing ground states\cite{Motta2020}. Starting from a reference $|\Phi\rang$, the state $e^{\tau\hat H}|\Phi\rang$ converges to the ground state of the Hamiltonian $\hat H$ in the $\tau\to-\infty$ limit, so long as $|\Phi\rangle$ has a non-vanishing overlap with the ground state. In practice, the propagation often requires representing and approximating the evolved state, for example by tensor networks\cite{White1992, ChanGKL2011, Chan2016, Cirac2021} or via sampling\cite{Ceperley1980, Reynolds1982, Foulkes2001, Booth2009, Shepherd2012, Wouters2014, Motta2018} in quantum Monte Carlo. Another possible ansatz is the coupled-cluster exponential\cite{Shavitt2009} $e^{\hat T(\tau)}|\Phi\rang$, which has been used in finite temperature coupled-cluster theory\cite{Sanyal1992, White2018, White2020} to express the thermal state $e^{\beta\hat H}$, and in moment coupled-cluster theory\cite{Ai2025} to represent the moment generating function $\lang e^{\tau\hat H}\rang$. This ansatz has also been studied in several real-time extensions of coupled-cluster theory\cite{Monkhorst1977, Dalgaard1983, Takahashi1986, SverdrupOfstad2023}. Here, we discuss how to use a coupled-cluster ansatz to represent the imaginary-time evolution trajectory from an arbitrary state. Starting with single determinant initial states, we study the properties of the evolution trajectories produced by this formalism, and discuss their connections to the standard coupled-cluster amplitude equations. We then introduce a coupled-cluster energy variance which can be computed from the evolution trajectory, and show how this allows us to obtain additional coupled-cluster solutions from trajectories in regimes where the standard amplitude equations are hard to solve or produce unphysical results. Finally, we generalize the above-mentioned discussion to arbitrary references where we find that the imaginary-time evolution provides a robust numerical solution strategy.

\section{Cluster ansatz for Imaginary-time evolution}
\label{sec:2}

Starting from an arbitrary reference state $|\Phi\rang$, the imaginary-time evolved state is $e^{\tau\hat H}|\Phi\rang$. For simplicity, we assume the Hamiltonian is normal-ordered\cite{Kutzelnigg1997} w.r.t. $|\Phi\rang$. We may write down a cluster ansatz for the evolved state as $e^{\tau\hat H}|\Phi\rang = e^{\hat T(\tau)}|\Phi\rang$, where the (generalized) cluster excitation operator reads\cite{Kutzelnigg1997}
\begin{equation}
\begin{aligned}
    \hat T(\tau) &= T_0(\tau) + \sum_{AI} T_{AI}(\tau) \{c_A^\dagger c_I\} \\
    &+ \frac{1}{(2!)^2}\sum_{ABIJ} T_{ABIJ}(\tau) \{c_A^\dagger c_B^\dagger c_Jc_I\} + \cdots 
\end{aligned}
\end{equation}
where $c^\dagger/c$ are electron creation/annihilation operators, and $A/I$ refer to generalized virtual/occupied orbitals. We note that a scalar term $T_0(\tau)$ is needed to track the norm of $e^{\tau\hat H}|\phi\rang$ since this state does not follow the intermediate normalization $\lang\phi |\psi\rang = 1$ as in standard coupled-cluster theory. When $|\Phi\rang$ is a single-determinant $|\phi\rang$, we shall use $a/i$ instead of $A/I$. We start with a single-reference $|\phi\rang$ to discuss the basics of the theory. 
\begin{figure*}[!ht]
    \centering
    \includegraphics[width=0.50\linewidth]{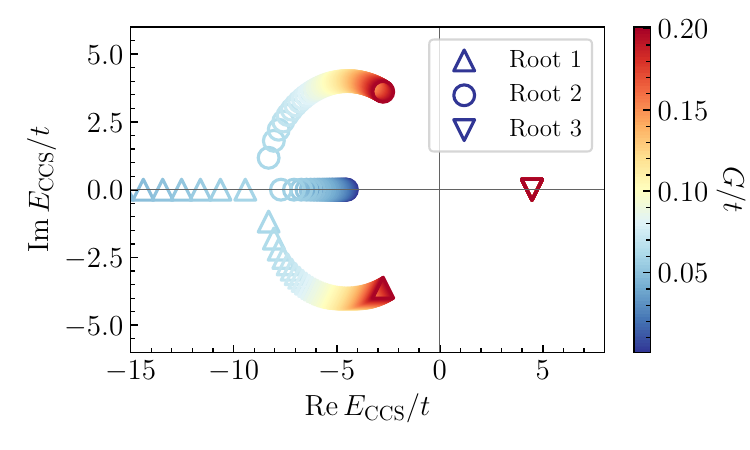}
    \includegraphics[width=0.40\linewidth]{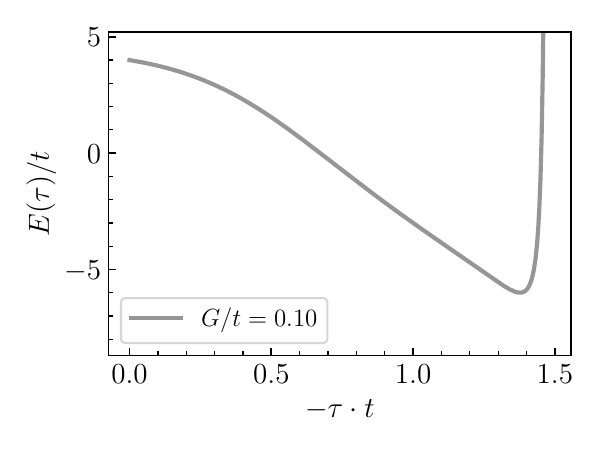}

    \caption{Left panel: Coupled-cluster singles energies in the complex plane corresponding to different roots of the cubic amplitude equation. The system is a Hubbard dimer with hopping $t$, fixed on-site interaction $U/t=4$, and variable pair hopping $G$. When $G/t\to0$, the energy of root 1 approaches $E\to-\infty$. Energy of root 3 is insensitive to $G/t$. Right panel: Energy of the imaginary-time evolved wavefunction $E(\tau)$ under the coupled-cluster singles approximation at $G/t=0.10$.}
    \label{fig:00-sr-hubdim}
\end{figure*}

We note that $e^{\hat T(0)} = \hat 1$. From $\tau=0$, we may evolve $|\phi\rang$ in imaginary time, 
and from the relation $e^{(\tau + d\tau)\hat{H}}|\phi\rangle = e^{\hat T(\tau + d\tau)}|\phi\rang$, we obtain the following differential equation for $\hat T(\tau)$\cite{SverdrupOfstad2023}:
\begin{equation}
    e^{-\hat T(\tau)}\hat He^{\hat T(\tau)}|\phi\rang =\frac{\partial \hat T(\tau)}{\partial \tau} |\phi\rang
    \label{eqn:ite-cc-sr}
\end{equation}

This differential equation formally gives the exact imaginary-time evolution state trajectory as $e^{\hat T(\tau)}|\phi\rang$ when $\hat T(\tau)$ is not truncated. A natural consequence is that exact imaginary-time evolution and standard full coupled-cluster theory should find the same exact ground state. Since the exact ground state has well-defined and finite amplitudes in Fock space for all $\tau$ (up to a scalar normalization factor), the non-scalar amplitudes $\hat T'(\tau) = \hat T(\tau) - T_0(\tau)$ should be finite in the $\tau\to-\infty$ limit. For such a limit to exist, the derivatives $\partial_\tau \hat T'(\tau)$ must vanish at $\tau\to-\infty$, and from equation (\ref{eqn:ite-cc-sr}), the amplitudes must satisfy the standard coupled-cluster amplitude equations $(1-|\phi\rang\lang \phi|) e^{-\hat T(-\infty)}\hat He^{\hat T(-\infty)}|\phi\rang = 0$. Therefore, with the full cluster ansatz, a $\tau\to-\infty$ limit always exists and the imaginary-time evolution trajectory coincides with a solution of the classical coupled-cluster amplitude equations in that limit.

While this connection is clear for exact theories, when $\hat T(\tau)$ is truncated (for example, we may set $\hat T(\tau)$ to only couple $|\phi\rang$ with a smaller excitation space $\hat Q \subset 1-|\phi\rang\lang \phi|$), the corresponding differential equation (\ref{eqn:ite-cc-sr}) projected to the excitation space no longer gives exact imaginary-time evolution trajectories. The characteristics of these approximate trajectories are less known and worth looking into. Following the argument above, if the non-scalar amplitudes $\hat T'(\tau)$ are still finite at $\tau\to-\infty$, the corresponding derivatives must again vanish, and $\hat T'(-\infty)$ must still satisfy the truncated standard CC amplitude equations. However, an approximate trajectory can also break this connection, as there is no guarantee that a finite $\hat T'(-\infty)$ must exist. 

For example, we consider the simple (extended) Hubbard dimer model 
\begin{equation}
\begin{aligned}
    \hat H &= -\sum_\sigma t (c_{1\sigma}^\dagger c_{2\sigma} + c_{2\sigma}^\dagger c_{1\sigma}) + U(n_{1\uparrow} n_{1\downarrow} + n_{2\uparrow} n_{2\downarrow}) \\ &+ G(c_{2\uparrow}^\dagger c_{2\downarrow}^\dagger c_{1\downarrow} c_{1\uparrow} + \mathrm{h.c.})
\end{aligned}
\end{equation}
with $U/t = 4$ and we start from a reference $|\phi\rang = |\uparrow\downarrow, 0\rang$, where the first site is doubly occupied and second site is empty. We then solve the spin-free coupled-cluster singles (CCS) equation with cluster operator $\hat T_1 = T(c_{2\uparrow}^\dagger c_{1\uparrow} + c_{2\downarrow}^\dagger c_{1\downarrow})$. The standard amplitude equation is given by
\begin{equation}
    -GT^3 + tT^2 - UT - t = 0
    \label{eqn:00-hubdim-ccs}
\end{equation}
and has three roots when $G\neq 0$. When $G/t < 0.06$ and all solutions of (\ref{eqn:00-hubdim-ccs}) are real, imaginary-time evolution with CCS converges to root 2 in the $\tau\to-\infty$ limit. From Figure~\ref{fig:00-sr-hubdim}, when $G/t$ decreases, root 1 moves toward $E=-\infty$ and root 3 barely responds to the change of $G$, always having positive correlation energy. Therefore, root 2 is a more reasonable approximation of the ground state, and it is reasonable that imaginary-time evolution converges to it. However, we note that equation (\ref{eqn:00-hubdim-ccs}) can have complex solutions when the pair hopping $G/t>0.06$, and the CCS energy $E_\mathrm{CCS} = U+T(GT - 2t)$ also can become complex, shown in the left panel of Figure~\ref{fig:00-sr-hubdim}, while imaginary-time evolution following (\ref{eqn:ite-cc-sr}) always produces real amplitudes and real energies as long as the Hamiltonian is real. When $G/t$ grows and root 2 gives a complex energy, it is impossible for ITE to converge to it. From the right panel of Figure~\ref{fig:00-sr-hubdim}, at $G/t=0.10$, imaginary-time evolution diverges as $E(\tau)$ goes to $+\infty$ already at finite $\tau$, and the trajectory no longer connects to root 2, which is now complex. 

Intuitively, the higher powers of the Hamiltonian in $e^{\tau\hat H}$ tend to contribute more at larger $\tau$. This introduces a larger number of connected diagrams in the underlying many-body representation of the energy\cite{Ai2025}. For a coupled-cluster ansatz with a fixed number of connected skeletons (as is the case with a truncated coupled-cluster ansatz), it is expected that the quality of the approximation drops as $\tau$ grows, which could cause the evolution to diverge, if the ground state is not reached quickly enough. In the Hubbard dimer context, the CCS ansatz seems to be poor in quality for large $G/t$ as it gives unphysical complex energies as the solution of the amplitude equation, and imaginary-time evolution also diverges. The lack of a $\tau\to-\infty$ limit such as this is not uncommon for approximate evolutions in practice.

Fortunately, when the CC amplitude equations have no reasonable solutions, we can still extract useful information from the imaginary-time trajectories before they diverge. Intuitively, at shorter times, imaginary-time evolution should be more reliable as there are fewer missing connected pieces in the energy $E(\tau)$. For example, in the right panel of Figure~\ref{fig:00-sr-hubdim}, before it reaches a local minimum and subsequently diverges, the energy $E(\tau)$ does improve from the reference energy at the start of the evolution. Numerically, we may use the energy variance\cite{Ye2017} 
\begin{equation}
    \sigma(\tau) = \frac{\lang \phi|\hat H^2e^{\tau\hat H}|\phi\rang}{\lang \phi|e^{\tau\hat H}|\phi\rang} - \frac{\lang \phi|\hat He^{\tau\hat H}|\phi\rang^2}{\lang \phi|e^{\tau\hat H}|\phi\rang^2}
\end{equation}
to characterize how close the imaginary-time evolved wavefunction is to an eigenstate of $\hat H$ at a given time $\tau$. Within our formalism, since the projected expression for the energy is $E(\tau) = \lang \phi|\hat He^{\tau\hat H}|\phi\rang/\lang \phi|e^{\tau\hat H}|\phi\rang$, the variance can be computed in a way that is consistent with the underlying approximation to the energy as $\sigma(\tau)=\partial_\tau E(\tau)$, which we refer to as the coupled-cluster energy variance. Note that the coupled-cluster energy variance only coincides with the exact energy variance when there is no truncation, as in a truncated theory, $E(\tau)$ is also computed approximately. In the following, when we refer to the variance $\sigma(\tau)$, we will always mean the coupled-cluster energy variance, defined in this way. In the Hubbard dimer example, since $E(\tau)$ eventually increases with $-\tau$, there is a point where the variance $\sigma(\tau)$ defined as $\partial_\tau E(\tau)$
becomes negative, which further hints at the poor quality of the given CC approximation since $\sigma(\tau)\geq 0$ always holds for the exact theory. It is thus reasonable to take the point with the lowest variance [i.e., lowest non-negative $\sigma(\tau)$] to serve as one of the ``best'' estimates we can extract from the diverging trajectory. We note that since the coupled-cluster variance itself is approximate, its local minimum cannot be rigorously guaranteed to give the best estimate of the exact properties of the system, but we can consider this solution condition to be in the same spirit of ``best'' as the usual projected amplitude equations, namely it is satisfied by the exact solution, and may be applied when only information at a truncated CC level is accessible. In Sec.~\ref{sec:3}, with more realistic numerical examples, we discuss both cases (where the $\tau\to-\infty$ limit exists or not) in more detail, and for the latter case, we demonstrate how the  variance $\sigma(\tau)$ helps us extract the best possible estimates of ground-state observables in practice. 

\section{Properties of approximate trajectories}
\label{sec:3}

\begin{figure}[!ht]
    \centering
    \includegraphics[width=0.7\linewidth]{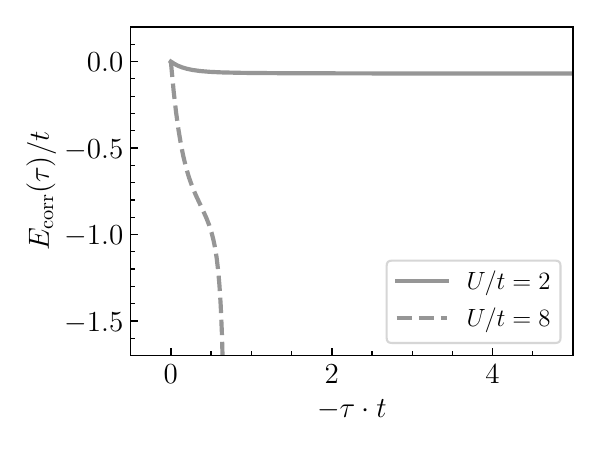}
    \includegraphics[width=0.7\linewidth]{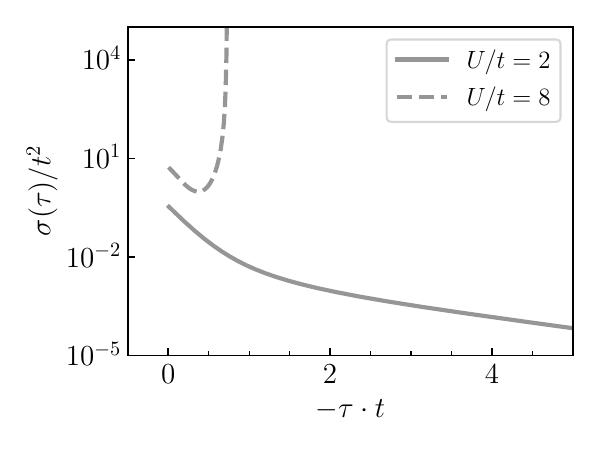}

    \caption{Correlation energy (upper panel) and  variance (lower panel) of the imaginary-time evolution trajectories under the CCSD approximation for a 30-site Hubbard chain at different $U/t$.}
    \label{fig:03-sr-hub}
\end{figure}

\begin{figure}[!ht]
    \centering
    \includegraphics[width=0.8\linewidth]{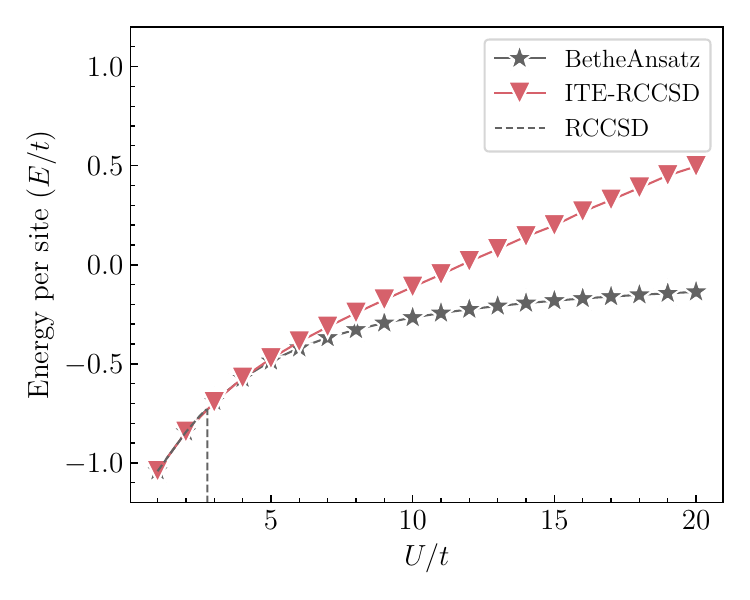}
    \includegraphics[width=0.8\linewidth]{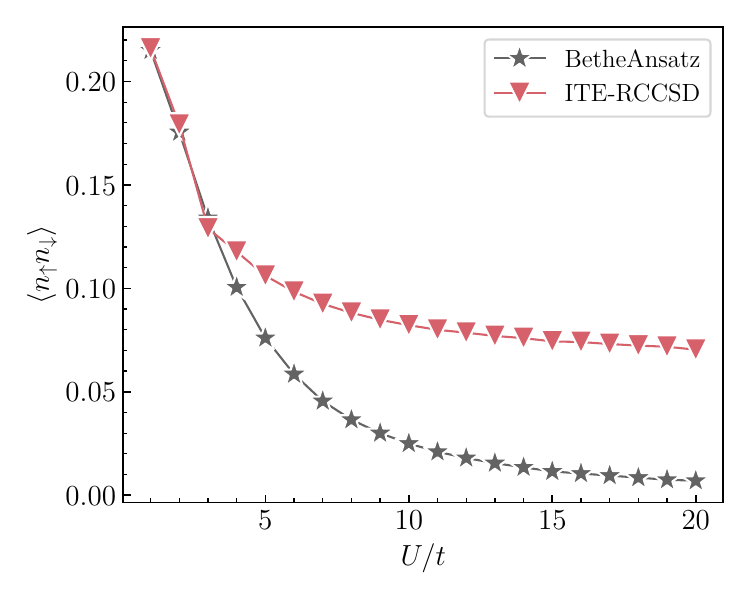}

    \caption{Energy (upper panel) and average double occupancies (lower panel) per site at the ITE variance minimum under the CCSD approximation for a 30-site Hubbard chain at different $U/t$.}
    \label{fig:04-sr-hub}
\end{figure}

The standard CCSD/density matrix renormalization group (DMRG) numerical benchmarks in this work were generated with the PySCF\cite{Sun2015, Sun2018, Sun2020}/block2\cite{Zhai2021,Zhai2023} software packages. We first study a half-filled 30-site 1D Hubbard chain, generalizing the Hubbard dimer example in Sec.~\ref{sec:2}. We run ITE using the coupled-cluster singles and doubles (ITE-CCSD) ansatz starting from a spin-restricted Hartree-Fock (RHF) reference and then track the energy $E(\tau)$ and variance $\sigma(\tau)$ along the propagation. The resulting trajectories at different interaction strengths $U/t$ are given in Figure~\ref{fig:03-sr-hub}. From the upper panel, we observe that at small $U/t=2.0$, ITE-CCSD converges quickly and gives a well-defined energy in the numerical $\tau\to-\infty$ limit, while at larger $U/t = 8.0$, the energy diverges to $-\infty$ and a finite $\tau\to-\infty$ limit is not found. We may compare to the standard CCSD energy references shown in the upper panel of Figure~\ref{fig:04-sr-hub}. For this system, CCSD energies tend to drop sharply at around $U/t = 2.76$, after which the amplitude equations also become very difficult to converge. We note that similar trends have been reported in shorter Hubbard chains\cite{Bulik2015}. In agreement with the previous general discussion, at $U/t=2.0$ where the $\tau\to-\infty$ limit exists, ITE-CCSD gives a solution to the CCSD amplitude equations, and they converge to the same energy estimate. In addition, at $U/t=8.0$, the limit does not exist for ITE-CCSD and relatedly, it is also very difficult to find a real solution of the standard amplitude equations numerically. 


From the lower panel of Figure~\ref{fig:03-sr-hub}, we note that at $U/t=2.0$, the variance $\sigma(\tau)$ takes the lowest possible non-negative value (which is zero in this case) also in the $\tau\to-\infty$ limit, where it corresponds to a standard CC solution. 
At larger $U/t=8.0$, although ITE-CCSD does not have a $\tau\to-\infty$ limit, there is still a well-defined point at finite $\tau$ where $\sigma(\tau)$ takes the lowest positive value during the propagation. 
Accordingly, we think of the imaginary-time evolved wavefunction at this variance minimum as the best estimate we may extract from the diverging trajectory. A variance minimum such as this (either at finite or infinite $\tau$) exists for the whole range of $U/t=1\sim 20$ that we have studied and, unlike the solutions to the standard amplitude equations, they are always easily found by the ITE process numerically. In the upper panel of Figure~\ref{fig:04-sr-hub}, we show the energy estimates at the variance minima across different $U/t$. Comparing with the exact result, while the ITE-CCSD energies eventually fail to capture the correct atomic limit at $U/t\to\infty$, they are still accurate to $0.03t$ up to $U/t=6$, where the standard CCSD amplitude equations are already hard to converge. 

We further compute the average double occupancies per site at the variance minima to provide 
insight related to the incorrect atomic limit. We note that at finite $\tau$, the amplitudes from ITE are usually not solutions to the standard amplitude equations, and a generalization of the standard $\Lambda$ equation to finite $\tau$ is desired to compute properties such as density matrices. To compute properties such as density matrices, we may further introduce a cluster ansatz for the imaginary-time evolved conjugate state $\lang \phi|e^{\tau\hat H} = \lang\phi|\hat \Lambda(\tau) e^{-\hat T(\tau)}$, where $\hat \Lambda(\tau)$ is a de-excitation operator and in turn satisfies
\begin{equation}
    \lang \phi|\hat \Lambda(\tau)e^{-\hat T(\tau)}\hat He^{\hat T(\tau)}+\lang \phi|\hat \Lambda(\tau)\frac{\partial \hat T(\tau)}{\partial \tau } = \lang \phi|\frac{\partial \hat \Lambda(\tau)}{\partial \tau}
\end{equation}

If a $\tau\to-\infty$ limit exists, $\hat \Lambda(-\infty)$ divided by its scalar part $\Lambda_0(-\infty)$ satisfies the standard CC $\Lambda$ equations\cite{Shavitt2009}. The definition then provides a consistent way to compute expectation values other than the energy at finite $\tau$. As we observe from the lower panel of Figure~\ref{fig:04-sr-hub}, the double-occupancies computed from ITE-CCSD at smaller $U/t$ agree quite well with the Bethe ansatz\cite{Lieb1968, Shiba1972, Knizia2012} and the standard CCSD reference. However, it is also clear that the double occupancy predicted does not converge to the correct atomic limit, where it should vanish completely. Therefore, as a caveat, while the variance minima in ITE-CC do help in giving in some sense the best possible estimates when the corresponding standard truncated CC diverges, they do not completely fix the problems arising from a poor reference or the truncated CC ansatz\cite{Bartlett2006, Kats2013, Bulik2015}. In fact, we might interpret this minimum variance solution as describing a metallic state at finite energy above the ground-state, rather than the insulating phase of the true ground-state.

\begin{figure}[!ht]
    \centering
    \includegraphics[width=0.7\linewidth]{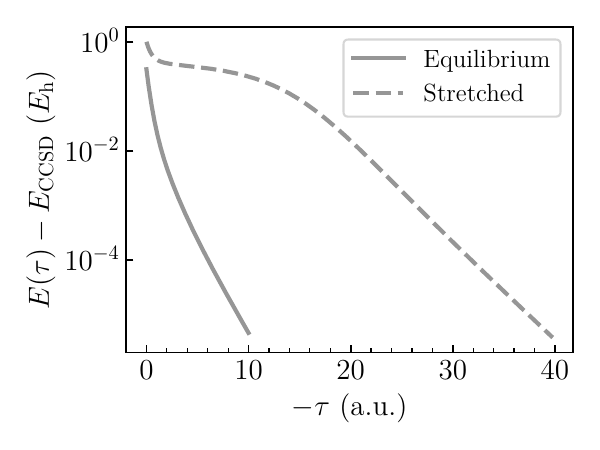}
    \includegraphics[width=0.7\linewidth]{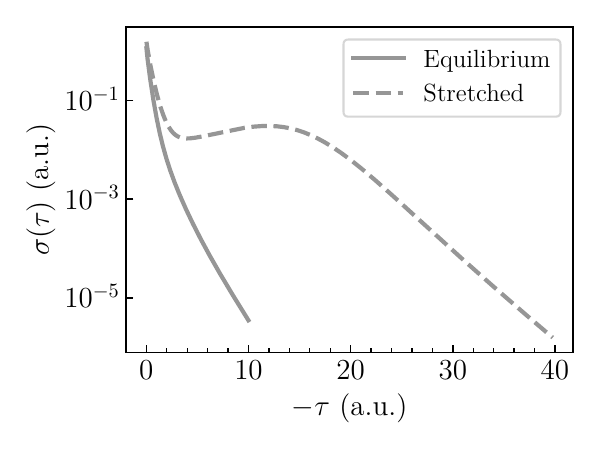}

    \caption{Energy difference from coupled-cluster references (upper panel) and variance (lower panel) of the imaginary-time evolution trajectories under the CCSD approximation for $\rm N_2$ at different geometries (cc-pVDZ basis).}
    \label{fig:01-sr-n2}
\end{figure}

We next move to the nitrogen dimer in the cc-pVDZ basis\cite{Dunning1989} as a molecular example. Again, we study ITE-CCSD starting from a RHF reference, and track the energy $E(\tau)$ and variance $\sigma(\tau)$. The trajectories at two typical bond lengths $R = 2.1 a_0$ (equilibrium) and $R = 5.1a_0$ (stretched) are shown in Figure~\ref{fig:01-sr-n2}. From the upper panel, we observe that a $\tau\to-\infty$ limit exists for both bond lengths and ITE-CCSD eventually converges to a CCSD solution with the same energy. However, although the minimal non-negative values of the variance $\sigma(\tau)$ appear at $\sigma(-\infty) = 0$, at the stretched geometry, there is also a local minimum of $\sigma(\tau)$ at finite $\tau$. 

\begin{figure}[!ht]
    \centering
    \includegraphics[width=0.9\linewidth]{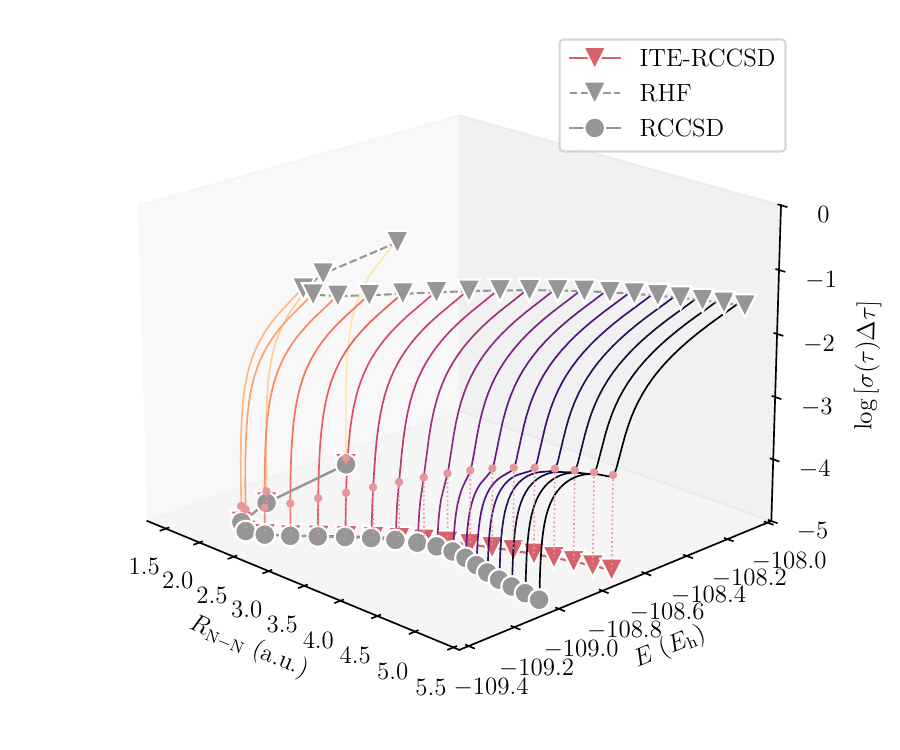}
    \includegraphics[width=0.8\linewidth]{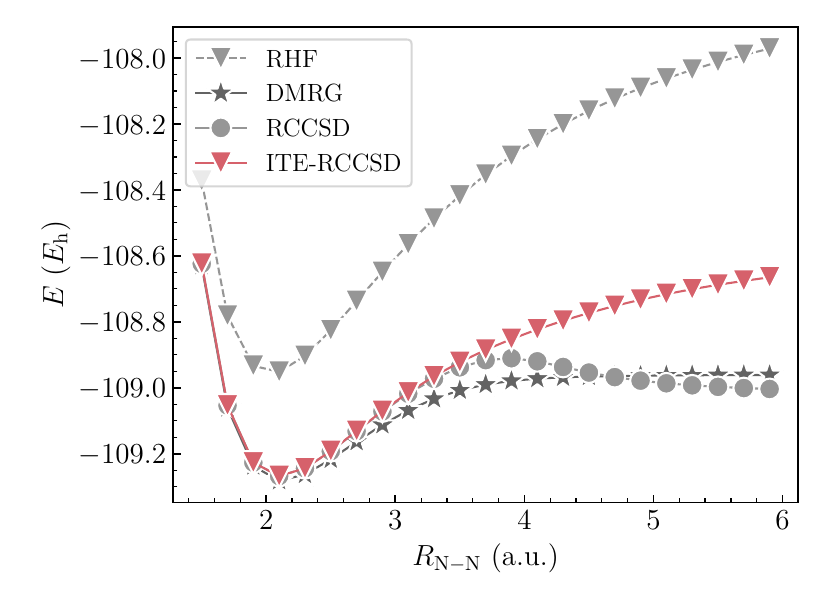}

    \caption{Upper panel: Imaginary-time evolution trajectories visualized in the space of bond length, energy, and variance for $\rm N_2$ (cc-pVDZ basis). $\Delta\tau = 0.02$ is the propagation step. The red dots denote the position of the transition line. Lower panel: Corresponding potential energy surfaces of the trajectories projected to the $R_\mathrm{N-N}-E$ plane. The red triangles correspond to the projection of the transition line.}
    \label{fig:02-sr-n2-var}
\end{figure}

It is interesting to examine these local minima and we show the $\sigma(E)$ trajectories at more bond lengths in Figure~\ref{fig:02-sr-n2-var}. For compactness, we plot against $E$ and since $E(\tau)$ is always a monotonically decreasing function of $-\tau$ here, a minimum in $\sigma(\tau)$ is also a minimum in $\sigma(E)$. From the upper panel, additional local minima of $\sigma(E)$ gradually appear as the bond length $R_\mathrm{N-N}$ grows, forming a smooth transition line that separates the evolution manifold. We compare the energies at the transition points to the standard CCSD reference energies in the lower panel of Figure~\ref{fig:02-sr-n2-var}. We note that as the molecule dissociates while using a spin-restricted reference, the system becomes more and more strongly-correlated and standard RCCSD theory shows an infamous turnover in the energy\cite{Kats2013, Bulik2015}. The transition line defined from the extra variance minima in ITE-CCSD, although slightly worse in terms of the absolute energy error than the traditional RCCSD curve, is free of any turnovers. Following the previous discussion, since the truncated cluster representation of the imaginary time evolution omits a larger number of connected contributions at large $\tau$, the variance minimum at the smallest $\tau$ value can be interpreted as suffering from the least amount of error accumulation, even though finite $\tau$ means that we have not yet projected to the ground-state energy scale. Indeed in this example, we observe that the transition line at smaller $\tau$ is able to remove the energy turnover, which is an unphysical behavior of the traditional curve.

All the examples suggest that for approximate ITE based on a truncated CC ansatz, it is useful to consider all the non-negative variance minima (global or local, finite or infinite $\tau$) as potential solutions. Based on this definition, there is always at least one solution for ITE-CC since $|\sigma(\tau)|$ is bounded from below and must have an infimum (minimum in practice) somewhere; when it appears at $\tau\to-\infty$, it coincides with a standard CC solution, and when it appears at finite $\tau$, it differs from the standard CC solution and can sometimes provide a more physically reasonable CC approximation. 

\begin{figure}[!ht]
    \centering
    \includegraphics[width=0.9\linewidth]{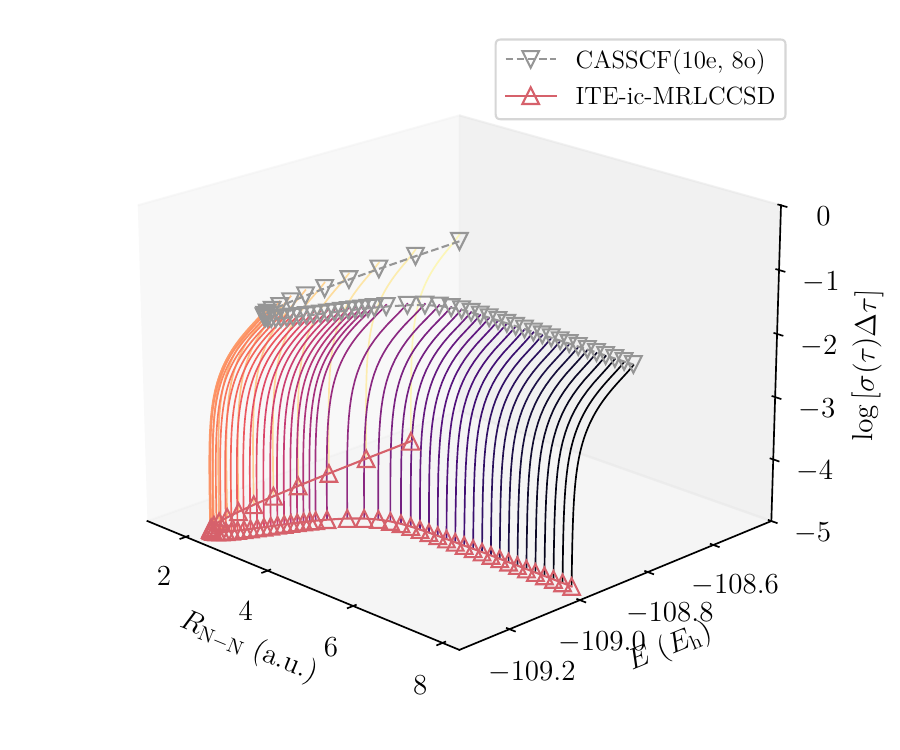}

    \caption{Multireference imaginary-time evolution trajectories under the linearized coupled-cluster approximation visualized in the space of bond length, energy and variance for $\rm N_2$ (cc-pVDZ basis). $\Delta\tau = 0.02$ is the propagation step.}
    \label{fig:05-mr-ic-var}
\end{figure}

\section{Time-Evolution from general references}

We have already demonstrated how a CC-like formalism describes imaginary-time evolution starting from a single determinant, $|\phi\rang$, and studied several properties of the evolution trajectories. We now investigate a more flexible framework that describes imaginary-time evolution from more general references, for example, a state $|\Phi\rang$ defined in a complete active space (CAS), as commonly used in strongly correlated systems.\cite{Roos1980, Olsen1988, Roos2016}

\begin{figure*}[!ht]
    \centering
    \includegraphics[width=0.54\linewidth]{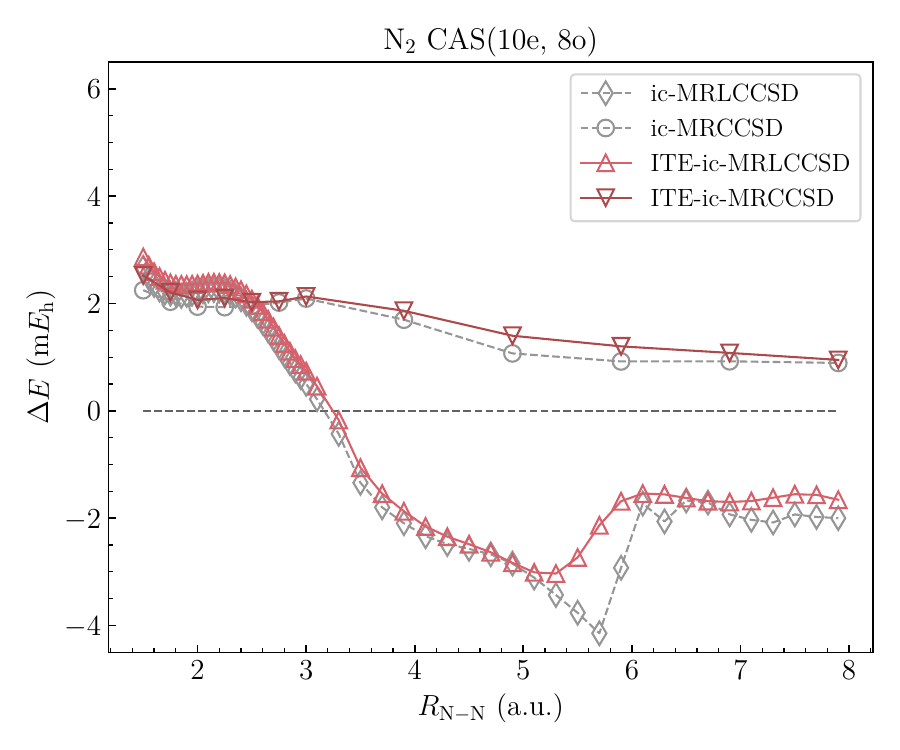}
    \includegraphics[width=0.36\linewidth]{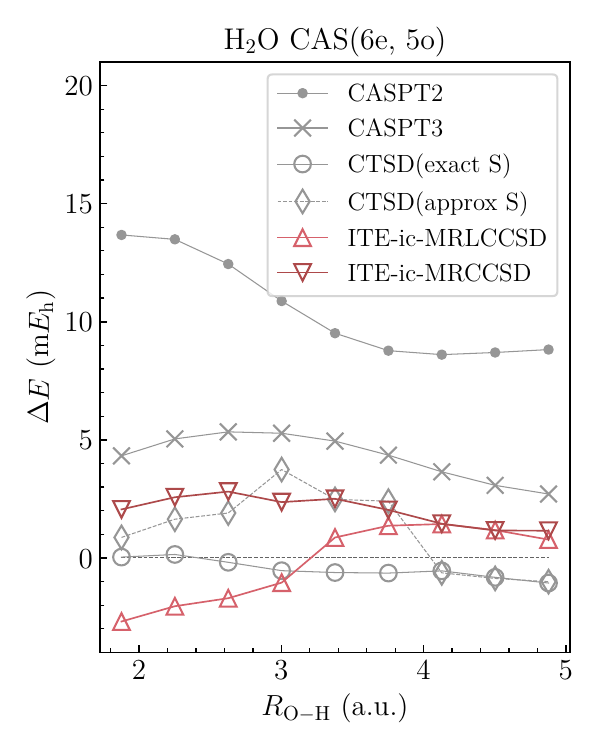}
    \caption{Energy error at the variance minimum found in multireference imaginary-time evolution under different CC approximations for $\rm N_2$ (left panel) and $\rm H_2O$ (right panel) (cc-pVDZ basis). For the ITE methods, we take DMRG with bond dimension $M=3000$ as the reference to compute $\Delta E$, and for the other standard methods, we take the reported full configuration interaction (FCI) energies in corresponding papers\cite{Black2019,Yanai2007} as reference.}
    \label{fig:06-mr-ic}
\end{figure*}

An extension based on generalized normal-ordering and Wick's theorem\cite{Kutzelnigg1997} is straightforward. We may write down the cluster ansatz $e^{\tau\hat H}|\Phi\rang = e^{\hat T(\tau)}|\Phi \rang$ with generalized indices, and the differential equation for $\hat T(\tau)$ that does the propagation reads
\begin{equation}
    e^{-\hat T(\tau)}\hat He^{\hat T(\tau)}|\Phi\rang = \left[\frac{1-e^{-\mathrm{ad}_{\hat T}}}{\mathrm{ad}_{\hat T}}(\partial_\tau \hat T)\right]|\Phi\rang
    \label{eqn:ite-cc-mr}
\end{equation}

where $\mathrm{ad}_{\hat T}(\hat X) = [\hat T,\hat X]_-$ and functions of $\mathrm{ad}_{\hat T}$ are given by their series expansions\cite{Hall2015}, for example, $(1-e^{-\mathrm{ad}_{\hat T}})/\mathrm{ad}_{\hat T} = 1-(1/2!)\mathrm{ad}_{\hat T}+(1/3!)\mathrm{ad}_{\hat T}^2-\cdots $. We note that nested commutators appear as $\hat T$ does not necessarily commute with its derivatives. Although truncations in $\hat T$ will be introduced to simplify Eq.~(\ref{eqn:ite-cc-mr}), it is still inconvenient to work with high-order commutators of $\hat T(\tau)$ and its derivatives. Here, we introduce a simplification by assuming that $[\hat T(\tau), \partial_\tau \hat T(\tau)]_-$ is small and we drop all terms involving commutators of them. Equation (\ref{eqn:ite-cc-mr}) then formally reduces to $e^{-\hat T(\tau)}\hat He^{\hat T(\tau)}|\Phi\rang = \partial_\tau \hat T |\Phi\rang$. We note that this simplification still gives the same $\tau\to -\infty$ limit if it exists because the derivative $\partial_\tau \hat T'(\tau)$ must again vanish and we recover the amplitude equation $e^{-\hat T(\tau)}\hat He^{\hat T(\tau)}|\Phi\rang=0$. Therefore, the simplification only has an effect on the path of imaginary-time evolution, which may change the variance minima if they exist at finite $\tau$, but leaves the $\tau\to-\infty$ limit intact. Under this simplification, we may simply project to the generalized excitation manifolds to solve for $\partial_\tau \hat T(\tau)$, and use this to propagate $\hat T(\tau)$,
\begin{equation}
    \lang \Phi_{IJ...}^{AB...}|e^{-\hat T(\tau)}\hat He^{\hat T(\tau)}|\Phi\rang = \lang \Phi_{IJ...}^{AB...}|\frac{\partial \hat T(\tau)}{\partial \tau} |\Phi\rang
    \label{eqn:ite-cc-mr-proj}
\end{equation}

We implemented the spin-free formalism of the above-mentioned internally-contracted imaginary-time evolution algorithm with automated code synthesis to generate the contractions\cite{Zhaitbp}. For a given $\tau$, $\partial \hat T/\partial\tau$ is obtained by solving the linear equation (\ref{eqn:ite-cc-mr-proj}) by the GMRES algorithm\cite{Saad1986, Sturler1999, Parks2006, Hicken2010} and used to propagate $\hat T(\tau)$. We test the implementation on two  molecules, $\rm N_\mathrm{2}$ and $\rm H_2O$, at various $R_\mathrm{N-N}$ and $R_\mathrm{O-H}$ bond lengths, starting from the corresponding spin-restricted CASSCF(10e, 8o) and CASSCF(6e, 5o) (complete active space self-consistent field) wavefunctions as references and run ITE using the internally-contracted multireference CCSD (ITE-ic-MRCCSD) ansatz\cite{Hanauer2011, Sharma2015, Aoto2016, Evangelista2018}. Following the standard ic-MRCCSD implementations\cite{Hanauer2011, Black2019}, we truncate the commutator expansion to the first order in linearized ITE-ic-MRCCSD and to the second order in ITE-ic-MRCCSD to remove the need for high-order density matrices. 

We first look at the features of the imaginary-time evolved trajectories. Interestingly, the trajectories from ITE-ic-MRCCSD and its linearized version are similar across all bond lengths. As an example, we show the $\sigma(E)$ curves for linearized ITE-ic-MRCCSD for $\rm N_2$ in Figure~\ref{fig:05-mr-ic-var}. We observe that there is only one variance minimum at $\tau\to-\infty$, and ITE approaches this limit smoothly without numerical problems. According to the previous discussion, ITE should give us the solution to the standard ic-MRCC amplitude equations, and we further verify this from the data shown in Figure~\ref{fig:06-mr-ic}. In the left panel, we compare ITE-ic-MRCC energies at $\tau\to-\infty$ to the reported corresponding standard ic-MRCC reference energies\cite{Black2019} for $\rm N_2$, and they agree well overall. We note that at around $R_\mathrm{N-N}=5.5a_0$, the reported linearized ic-MRCCSD energy shows a sudden jump, while the energy from the ITE $\tau\to-\infty$ limit is smoother. This difference is likely explained by the fact that internally-contracted methods are, in general, sensitive to the truncations made in orthogonalizing the excitation manifold. In the ITE solutions, we did not perform any explicit truncation of the manifold. These results again support the robustness of imaginary-time evolution as a numerical tool. In the right-panel of Figure~\ref{fig:06-mr-ic}, we demonstrate another application on $\rm H_2O$, where we study its symmetric dissociation. The variance minimum of ITE-ic-MRCC was again found numerically at $\tau\to-\infty$. The ITE-ic-MRCCSD provides a quite accurate dissociation curve, with a non-parallelity error of 1.65 $\mathrm{m}E_\mathrm{h}$, surpassing that of multireference perturbation theories such as CASPT2/CASPT3\cite{Andersson1990, Andersson1992, Werner1996, Celani2000}, and is comparable with that of earlier canonical transformation theory calculations\cite{White2002, Yanai2006, Yanai2007, Yanai2010, Neuscamman2010}.

\section{Summary}
We have discussed how to carry out imaginary-time ($\tau$) evolution from a given reference state within a coupled-cluster-like formalism, and studied the features of the trajectories that it produces under truncation of the coupled-cluster ansatz. We also introduced a coupled-cluster energy variance that is trivial to compute along the trajectory, which would coincide with the true energy variance when there is no truncation. Depending on the quality of the CC ansatz and the underlying reference, we find that a finite $\tau\to-\infty$ limit may or may not exist. When the limit exists, the trajectory converges to a solution of the standard CC amplitude equation, and when it does not, the points of minimum coupled-cluster energy variance provide the ``best'' estimates that we can extract from the diverging trajectory. We further show through numerical examples that even when the infinite time limit exists, the variance minima at short times may provide physically more meaningful regularization of the coupled-cluster predictions. Imaginary-time evolution itself also serves as a robust numerical tool to converge amplitude equations when the solution exists, as we observe in the multireference case studies. The imaginary-time evolution coupled-cluster framework thus provides a way to extend existing coupled-cluster approximations to challenging scenarios where physically reasonable predictions, or numerical solutions, have so far been hard to obtain.

Beyond the specific examples considered here, our work raises additional questions for study. The minimum variance solutions constitute another set of coupled cluster solutions, and analogous to the situation with symmetry broken coupled cluster solutions, may carry different physical content from the standard solutions. How to combine information from multiple solutions is a similar problem as how to combine information from different symmetry broken calculations. In addition, the minimum energy variance principle defined here is not limited to coupled cluster theory, but in fact can be defined for any diagrammatic theory. This is because the machinery of many-body theory, although usually employed to evaluate $E(-\infty)$, can also be used to obtain $E(\tau)$ at any $\tau$, and thus $\sigma(\tau)$ also. Extension of these minimum energy variance studies to methods other than coupled cluster theory is thus a natural future direction.

\section{Supplementary Information}

Please refer to the supplementary information file for implementation details of ITE-CC and additional remarks on size-consistency and possible relations to broken-symmetry methods (S.~Figs.~1--3).

\section{Acknowledgments}

This work was supported by the US Department of Energy, Office of Science, under Grant No. DE-SC0018140.

\bibliography{itecc.bib}

\end{document}